\documentclass[10pt,prl,aps,twocolumn,showpacs,superscriptaddress,floatfix]{revtex4}
\usepackage{graphicx}

\usepackage{amssymb}

\begin{document}

\title{Thermodynamics of a gas of deconfined bosonic spinons in two dimensions}

\author{Anders W. Sandvik}
\affiliation{Department of Physics, Boston University, 590 Commonwealth Avenue, Boston, Massachusetts 02215}

\author{Valeri N. Kotov}
\affiliation{Department of Physics, University of Vermont, 82 University Place, Burlington, Vermont 05405}

\author{Oleg P. Sushkov}
\affiliation{Department of Physics, Boston University, 590 Commonwealth Avenue, Boston, Massachusetts 02215}
\affiliation{School of Physics, University of New South Wales, Sydney 2052, Australia}

\begin{abstract}
We consider the quantum phase transition between a N\'eel antiferromagnet and a valence-bond solid (VBS) in a two-dimensional system of $S=1/2$ 
spins. Assuming that the excitations of the critical ground state are linearly dispersing deconfined spinons obeying Bose statistics, we derive 
expressions for the specific heat and the magnetic susceptibility at low temperature $T$. Comparing with quantum Monte Carlo results for the J-Q model, 
which is a candidate for a deconfined N\'eel--VBS transition, we find excellent agreement, including a previously noted logarithmic correction in
the susceptibility. In our treatment, this is a direct consequence of a confinement length scale $\Lambda \propto \xi^{1+a} \propto 1/T^{1+a}$, where 
$\xi$ is the correlation length and $a>0$ (with $a\approx 0.2$ in the model).
\end{abstract}

\date{\today}


\maketitle

The excitations of a quantum antiferromagnet are spin waves (magnons) carrying spin $S=1$. At a conventional quantum phase transition in which the 
antiferromagnetic (N\'eel) order parameter vanishes continuously \cite{chakravarty89,chubukov94}, the magnons remain 
well-defined elementary excitations even at the critical point. The gapped ``triplon'' excitations in the nonmagnetic phase also have $S=1$. 
A different scenario has also been proposed, in which the magnons of a two-dimensional system fractionalize into independent (deconfined) 
$S=1/2$ spinons at the phase transition, which was termed the {\it deconfined quantum critical} (DQC) point \cite{senthil04}. In the non-magnetic 
phase, which in this case is a valence-bond solid (VBS) with spontaneously broken lattice symmetries \cite{read89}, the spinons are confined into 
excitations which carry spin $S=1$ and $S=0$. This proposal is supported by quantum Monte Carlo (QMC) simulations of a ``J-Q'' model 
\cite{sandvik07,melko08,lou09,sandvik10,banerjee10} (an $S=1/2$ Heisenberg model including four-spin or higher-order terms), but objections have also been 
raised \cite{jiang08,kuklov08,isaev09}. The deconfinement scenario violates the long-held ``Landau rule'' according to which 
order--order transitions breaking unrelated symmetries should be of first order.

In this {\it Letter} we provide evidence for deconfinement in the J-Q model based on its thermodynamic properties. Using a phenomenological
ansatz of deconfined bosonic spinons with linear dispersion $\epsilon(k)=ck$ at temperature $T=0$, we derive expressions for the 
specific heat and the magnetic susceptibility at $T>0$. In addition to the spinon velocity $c$, these quantities depend on the length scale 
$\Lambda (T)$ within which spinons are confined at $T>0$. As a consequence of the $T=0$ DQC theory \cite{senthil04}, one may expect 
this length to diverge when $T \to 0$ as a power of the correlation length, $\Lambda \propto \xi^{1+a}$, where the anomalous exponent $a>0$ and the 
correlation length diverges as $\xi \propto 1/T^z$ with dynamic exponent $z=1$. The correlation length is defined in the standard way, 
using the antiferromagnetic or VBS correlations (which diverge in the same way at the DQC point). We will show that confinement on the
larger length scale when $a>0$ implies a logarithmic correction to the standard quantum-critical scaling $\chi \propto T$ of the susceptibility, 
which agrees with a recent QMC study of the J-Q model \cite{sandvik10}. The specific heat and the susceptibility obey the spinon-gas forms with 
a common value of $c$ and $a\approx 0.2$.

{\it Spinon gas}---Our assumption is that a system with couplings tuned to the $T=0$ quantum-critical point can be described as a gas of bosonic 
spinons with dispersion $\epsilon(k)=\sqrt{c^2k^2+\Delta^2(T)}$ at $T>0$. This dispersion is valid for magnons at a conventional O($3$) quantum 
phase transition between the N\'eel state and a disordered state (e.g., in dimerized Heisenberg models 
\cite{singh88,sandvik94,troyer96,shevchenko00,brenig06,wang06,wenzel09}), in which case the thermal ``gap'' $\Delta$ is related to the correlation 
length $\xi$ according to $\Delta \propto 1/\xi \propto T^z$, with $z=1$ \cite{chakravarty89,chubukov94}. At the DQC point $\xi \propto 1/T$ is 
also expected, but the spinon gap should be given by the larger confinement length, $\Delta \propto 1/\Lambda \propto T^{1+a}$. Effectively, the
gap is used as an infrared cut-off in momentum space.

The $T=0$ confinement exponent $a$ was previously estimated using QMC results for the finite-size scaling of the U(1)-Z$_{\rm 4}$ 
cross-over of the VBS order-parameter symmetry (a hallmark of the DQC theory \cite{senthil04}) of a variant of the J-Q model. The result 
was $a=0.20 \pm 0.05$. It is unclear whether the $T>0$ confinement exponent should be the same, however. Here we leave $a$ 
as a free parameter along with the velocity $c$ and write the gap as
\begin{equation}
\Delta = m_{1/2}T\left ({T}/{c}\right)^a,
\label{deltaspinon}
\end{equation}
where the  constant $m_{1/2}$ should be close to $1$. In the case of magnons, it is known that $\Delta=m_1T$, with the mean-field value 
$m_1\approx 0.96$ \cite{chubukov94} in good agreement with QMC calculations of observables (e.g., the magnetic susceptibility) which depend
on this constant \cite{chubukov94,shevchenko00}.

In a magnetic field $B$, the spinon level is split into
\begin{equation}
\epsilon_{\pm}(k)=\sqrt{c^2k^2+\Delta^2} \pm \mu B \equiv \epsilon(k)  \pm \mu B,
\label{disp}
\end{equation}
where $\mu=1/2$. This form with $\mu=1$ holds also for the two shifted magnon levels (with $\epsilon_0$ not shifted). In the CP$^1$ DQC
theory \cite{senthil04}, there are both spinons and anti-spinons, which contribute equally to thermodynamic properties. We take this into account 
with a factor $F=2$, while for magnons $F=1$. 

With the boson occupation number $n(\epsilon)=1/(e^{\epsilon/T}-1)$ the magnetization per lattice site for small $B$ is:
\begin{eqnarray}
M&=&\mu F
\int\left(\frac{1}{e^{\epsilon_{-}/T}-1}-\frac{1}{e^{\epsilon_{+}/T}-1}\right)\frac{d^2k}{(2\pi)^2}\nonumber\\
&=&\mu^2 F\frac{TB}{4\pi c^2}\int_{0}^{\infty}\frac{x dx}{\sinh^2  [\frac{1}{2}\sqrt{x^2 + (\Delta/T)^2}]}.
\label{M}
\end{eqnarray}
The integral can be computed exactly,
\begin{equation}
\int_{0}^{\infty}\frac{x dx}{\sinh^2(\frac{1}{2}\sqrt{x^2 + p^2})} = \frac{4p}{1-{\rm e}^{-p}}-4\ln({\rm e}^p-1),
\label{integp}
\end{equation}
where $p=\Delta/T$. For magnons at the usual O($3$) transition, $p=m_1 \approx 0.96$ and the susceptibility is \cite{chubukov94}
\begin{equation}
\chi_1 \approx  (1.0760/\pi c^2)T.
\label{xmagnon}
\end{equation}
For spinons, if there is indeed an anomalous exponent $a>0$ in Eq.~(\ref{deltaspinon}), 
then $\Delta/T \to 0$ as $T \to 0$ and we can use the expansion of (\ref{integp}) around $p=0$, giving
\begin{equation}
\chi_{1/2} =  \frac{T}{2\pi c^2} \left [1 + a\ln{\left ( \frac{c}{T} \right )} + \frac{1}{24}\left ( \frac{T}{c} \right)^{2a} \right ].
\label{xspinon}
\end{equation}
Here we have used $m_{1/2}=1$ in (\ref{deltaspinon}), and the next correction to $\chi_{1/2}/T$ is of order $(T/c)^{4a}$. The logarithmic 
correction is very interesting, as it was already identified in a recent QMC study of the J-Q model \cite{sandvik10}. 

The specific heat per site is
\begin{eqnarray}
C_{S}=(2S+1)F \int\epsilon(k)\frac{\partial n(\epsilon)}{\partial T}\frac{d^2k}{(2\pi)^2},
\label{EC}
\end{eqnarray}
which for $S=1/2$ leads to the low-$T$ behavior
\begin{eqnarray}
&& C_{1/2} = \frac{2T^2}{\pi c^2}\times \label{cspinon} \\ 
&&~~~~~ \left [ 6\zeta(3)-\left ( \frac{T}{c} \right)^{2a}
\left [\frac{3}{2}+a+a(1+a)\ln{\left ( \frac{c}{T}\right )} \right ] \right ],  \nonumber
\end{eqnarray}
where we have again used $m_{1/2}=1$ and $\zeta(3)\approx 1.20206$.  Note that the log-correction is multiplied by a power and is not as 
dramatic as in the susceptibility (\ref{xspinon}). For the O($3$) transition, $S=F=1$ in (\ref{EC}) gives, at low $T$ \cite{sachdev93,chubukov94},
\begin{equation}
C_1 = [36 \zeta(3)/5\pi c^2]T^2.
\label{cmagnon}
\end{equation}

Apart from the logarithms, the differences in the thermodynamics between spinons and magnons arise mainly from the degeneracy factors and 
$\mu$. The log correction to $\chi_{1/2}$ in (\ref{xspinon}) is significant, however, and should be a decisive fingerprint of 
the deconfined spinon gas.

{\it Quantum-critical models}---A promising model exhibiting a N\'eel--VBS transition is the J-Q model, which in its simplest form is 
defined by the hamiltonian \cite{sandvik07}
\begin{equation}
H = -J\sum_{\langle ij\rangle}C_{ij} - Q\sum_{\langle ijkl\rangle}C_{ij}C_{kl},
\label{ham}
\end{equation}
where $C_{ij}$ is a singlet projector; $C_{ij}=1/4-{\bf S}_i \cdot {\bf S}_j$. In the J (Heisenberg) term $ij$ are nearest neighbors on the 
square lattice, while in the Q term $ij$ and $kl$ form opposite edges of a $2\times 2$ plaquette. There is mounting evidence 
\cite{sandvik07,melko08,lou09,sandvik10,banerjee10} of a continuous $T=0$ transition in this system between a N\'eel state for $J/Q > (J/Q)_c$ 
and a VBS for $J/Q < (J/Q)_c$, with $(J/Q)_c \approx 0.045$ \cite{sandvik10}. 

\begin{figure}
\centerline{\includegraphics[width=7.5cm, clip]{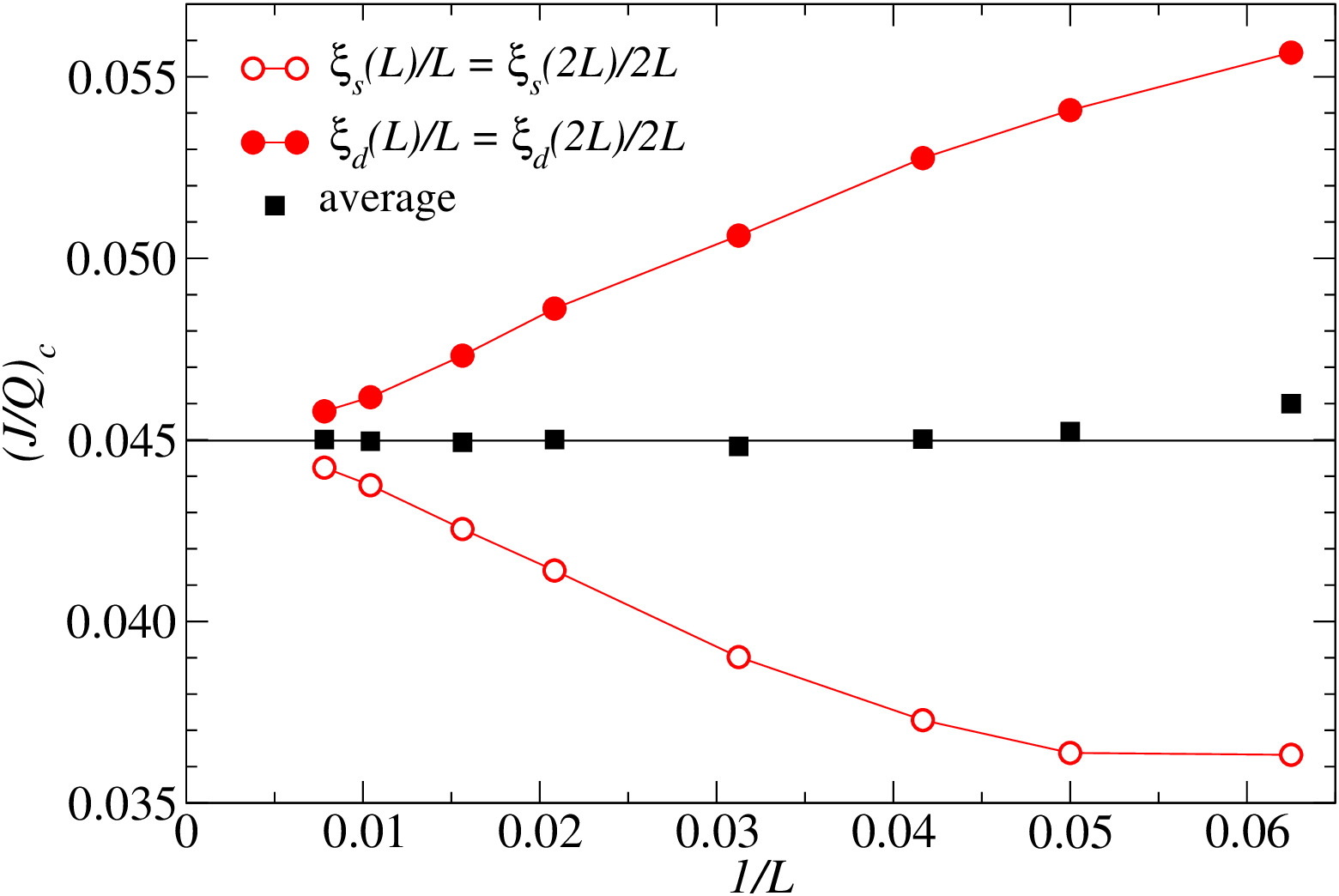}}
\vskip-2mm
\caption{(Color online) Size dependent critical coupling of the J-Q model based on $(L,2L)$ curve crossings of the scaled spin and dimer 
correlation lengths, $\xi_s/L$ and $\xi_d/L$. The average of these two estimates is also shown. The horizontal line shows the critical point 
extracted from the average.}
\label{jqc}
\vskip-3mm
\end{figure}

To ensure pure critical behavior when testing the above spinon gas predictions, it is necessary to know the critical coupling ratio $(J/Q)_c$ to high 
precision. Here we use the correlation lengths $\xi_s$ and $\xi_d$ extracted from, respectively, the spin-spin and dimer-dimer (four-spin) 
correlation functions. QMC calculations were carried out on $L\times L$ lattices with $L$ up to $256$ at inverse temperature $\beta=Q/T = L$ 
(as in \cite{sandvik10}). At a DQC point, $\xi_s/L$ and $\xi_d/L$ should be size independent for large $L$. Curves plotted versus $J/Q$ for two 
different system sizes, e.g., $L$ and $2L$, should then cross each other at some value $(J/Q)_L$, which can be different for $\xi_s/L$ and 
$\xi_d/L$ but in both cases should approach $(J/Q)_c$ when $L \to \infty$. Such crossing points are shown in Fig.~\ref{jqc}. Due to the slow 
convergence, it is difficult to extrapolate precisely. There is, however, a remarkable feature of these data: The $\xi_s/L$ and $\xi_d/L$ 
crossing points approach an apparent common asymptotic value at the same rate but from different sides. Their average exhibits almost no size 
dependence, and one can therefore obtain a much better critical-point estimate than what might initially have been expected. The result based 
on the four largest-$L$ points (which agree completely within statistical errors) is $(J/Q)_c=0.04498(3)$. Here we will use $J/Q=0.045$.

Conventional O($3$) $T>0$ scaling has been studied in the past in various dimerized Heisenberg models (where the hamiltonian 
itself breaks lattice symmetries, and no other symmetries are broken in the disordered phase) \cite{sandvik94,troyer96,shevchenko00,brenig06}. 
To compare with the J-Q model, we consider a system with couplings $J$ and $J'>J$, with the stronger ones arranged in columns. This 
model was the subject of a recent high-precision study \cite{wenzel09}, which gave the critical ratio $(J'/J)_c=1.9096(2)$. A further improved 
estimate is now available, $(J'/J)_c=1.90948(4)$ \cite{sandvikxx}. We use $J'/J=1.9095$ and carry out more detailed comparisons with the 
O($3$) theory than in past studies. In this case all quantities should be normalized per two-site unit cell.

\begin{figure}
\centerline{\includegraphics[width=7.2cm, clip]{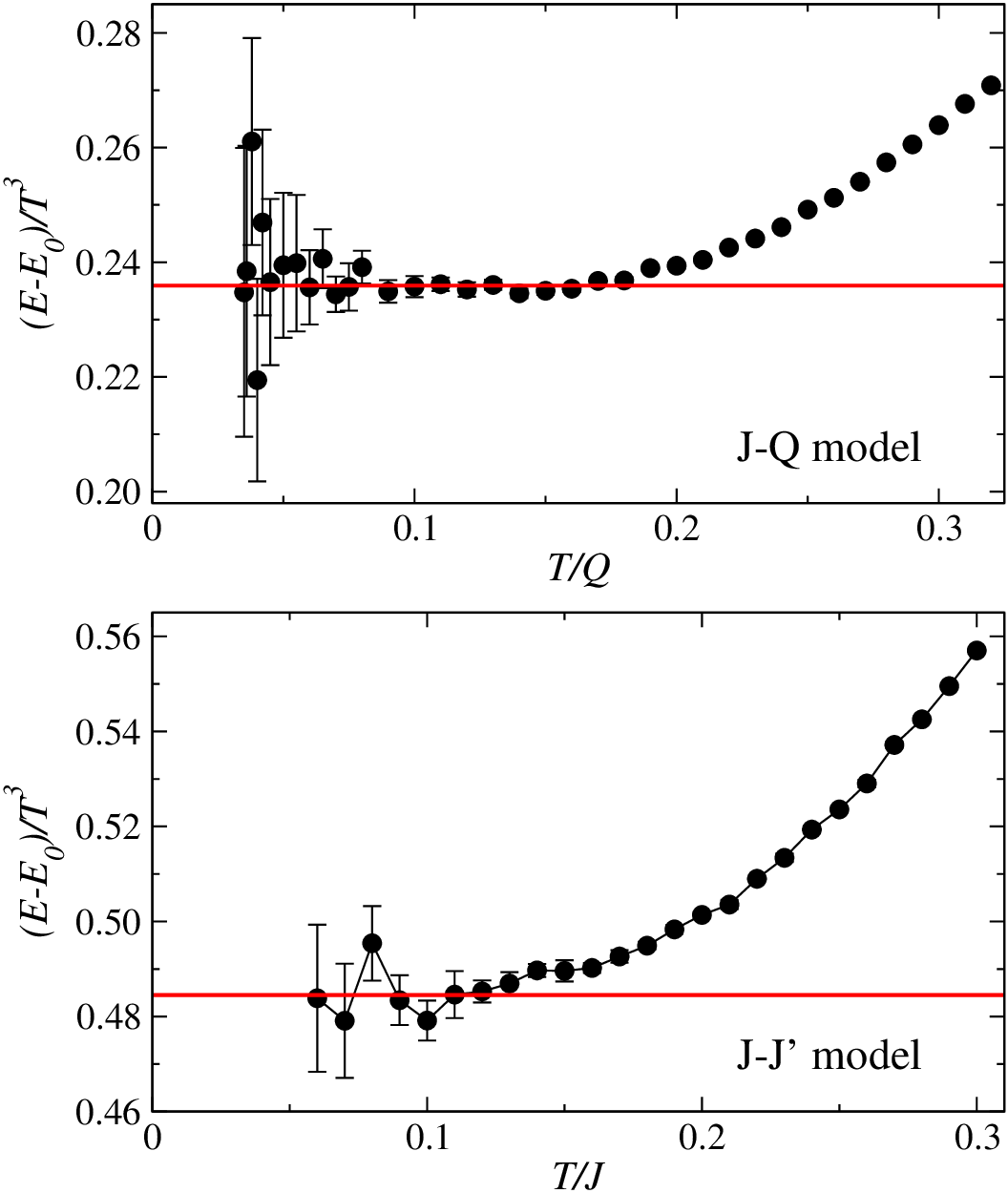}}
\vskip-2mm
\caption{(Color online) Temperature dependence of the energy $E$ relative to the ground state energy $E_0$ with the expected leading 
$T^3$ dependence divided out. The horizontal lines show the prefactor of the cubic term in the $E(T)$ fits.}
\label{efit}
\vskip-3mm
\end{figure}

{\it Energy fits}---We first fit low-$T$ results for the internal energy based on the leading specific heat forms (\ref{cspinon}) and 
(\ref{cmagnon}). QMC calculations were carried out using sufficiently large lattices ($L\le 512$) to eliminate finite-size effects 
in the range of temperatures considered. We also use the ground state energy extrapolated to $L=\infty$ based on $\beta=L$ results. 
For the J-Q model at $J/Q=0.045$, the result is $E_0/Q=-0.8740318(4)$, while the J-J' model with $J'/J=1.9095$ has $E_0/J=-1.740507(2)$ 
per unit cell. Fig.~\ref{efit} shows $E(T)$ after $E_0$ has been subtracted and $T^3$ has been divided out. The low-$T$ behavior gives 
the spinon velocity $c=2.55Q$ for the J-Q model and the magnon velocity $c=1.38J$ for the J-J' model (which should be interpreted as 
$c=\sqrt{c_xc_y}$ since the J-J' model is anisotropic). While there are corrections to the $T^3$ behavior in Fig.~\ref{efit}, the low-$T$ 
results for the J-Q model are not sufficiently accurate to test the correction in (\ref{cspinon}). It is anyway doubtful whether the 
spinon gas model can correctly capture subleading corrections. 

\begin{figure}
\centerline{\includegraphics[width=7.1cm, clip]{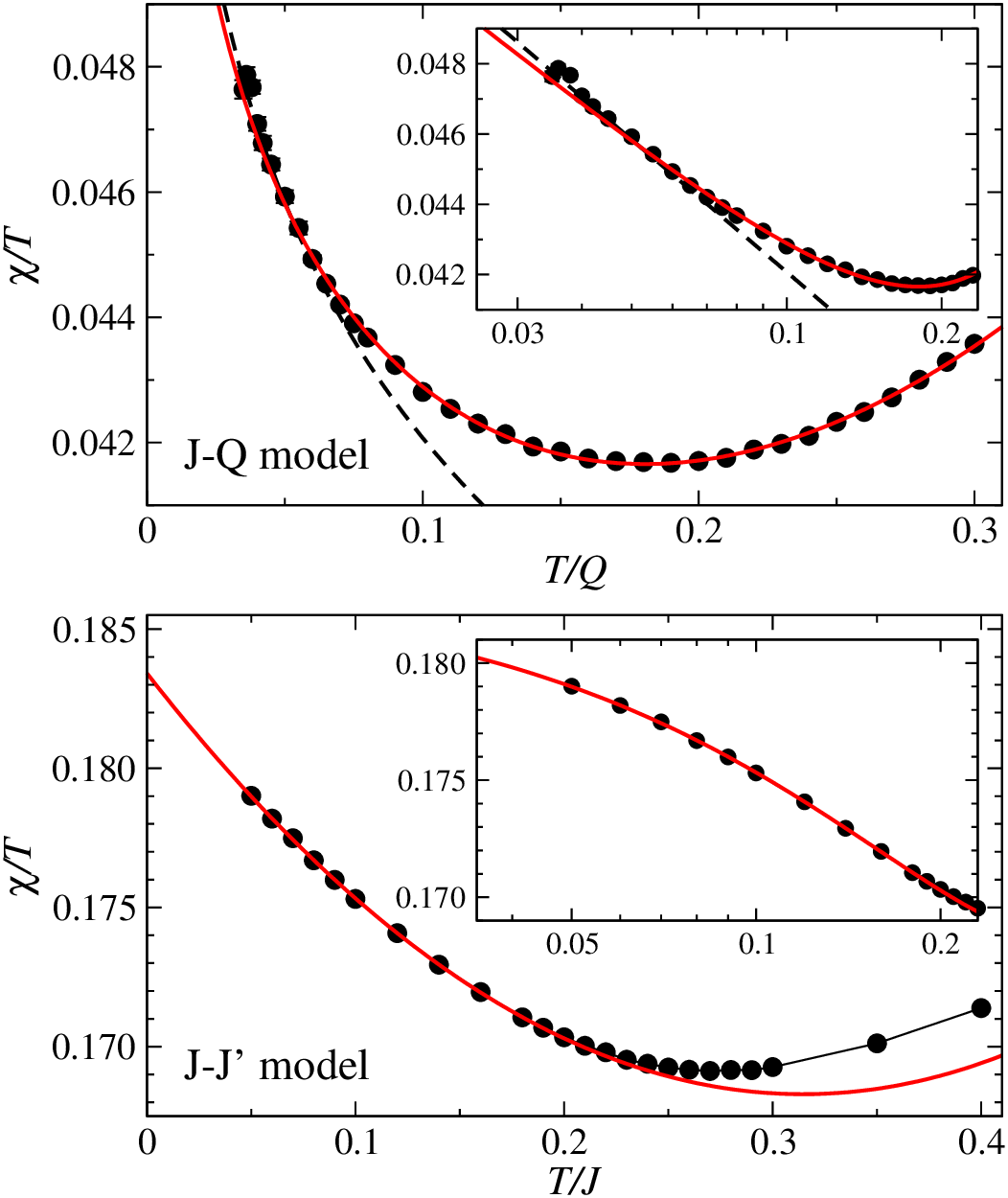}}
\vskip-2mm
\caption{(Color online) Susceptibility divided by $T$. The dashed curve in the J-Q graph is Eq.~(\ref{xspinon}) with $c=2.55Q$, $a=0.22$ 
(without the power-law correction). The solid curve has $c=2.48Q$, $a=0.20$, and also includes a $T^2$ term. The solid curve in the J-J' 
graph is a second-order polynomial. The insets show the same data on logarithmic $T$ scales.}
\label{suscfit}
\vskip-3mm
\end{figure}

{\it Susceptibility fits}---The susceptibilities of both models are shown in Fig.~\ref{suscfit} with $T$ divided out. In the case of the 
J-J' model, there are significant corrections to the asymptotic $T\to 0$ constant behavior expected with (\ref{xmagnon}). A second-order polynomial 
fit to the low-$T$ data is shown. The extrapolated $T=0$ susceptibility corresponds to a spin-wave velocity $c=1.36J$ in (\ref{xmagnon}), in
excellent agreement with the value obtained from the energy. 

Two different fits to J-Q results are shown in Fig.~\ref{suscfit}. Neglecting the power-law correction in (\ref{cspinon})
and fixing the spinon velocity to the value $c=2.55Q$ obtained from the energy, only the exponent $a$ in front of the log correction was 
adjusted to match the low-$T$ data. This result is $a=0.22$. A completely independent fit to a wider range of data, and including a 
$T^2$ correction (which was found to describe the data well in \cite{sandvik10}), gives $c=2.48Q$ and $a=0.20$. In both cases, including also
the very small power-law correction in (\ref{cspinon}) changes $\chi_{1/2}/T$ by less than $1\%$ and barely affects the extracted parameters. 

We have set $m_{1/2}=1$ in (\ref{deltaspinon}) throughout the above analysis, while we may only expect $m_{1/2} \approx 1$ [as with the constant 
$m_1$ in the conventional O($3$) theory]. Physical observables depend only weakly on $m_{1/2}$, however, and the consistent $c$-values extracted 
from two different quantities justify the use of $m_{1/2}=1$ {\it a posteriori}. 

{\it Wilson ratio}---The Wilson ratio of the J-Q model exhibits a log divergence. From Eqs.~(\ref{xspinon}) 
and (\ref{cspinon}) and the parameters of the fits (taking $c$ from the energy fit and $a$ from the susceptibility fit with $c$ fixed) we get 
$W_{1/2}=\chi T/C=w_{1/2}[1+ a\ln({c}/{T})]$, with $w_{1/2}=0.0346 \pm 0.0002$, $c=2.55 \pm 0.02$, and $a=0.222 \pm 0.005$. For the J-J' model we get 
$W_1=0.1262 \pm 0.0006$, in good agreement with $W_1=0.1243$ from Eqs.~(\ref{xmagnon}) and (\ref{cmagnon}). Including the next term in the $1/N$ 
expansion of the large-$N$ O($3$) theory \cite{chubukov94} makes this agreement worse by several percent, however. Note that if the log-correction is 
disregarded, $W_{1/2}$ is only about $1/4$ of $W_1$.

{\it Conclusions and discussion}---We have tested a model of non-interacting (deconfined) bosonic spinons against QMC data for the 
J-Q model, which is a promising candidate for a DQC point. The most notable result is that a confinement length $\Lambda$ diverging 
as $1/T^{1+a}$ with $a>0$ leads to a logarithmic correction to the susceptibility $\chi$, as was previously observed in 
the J-Q model \cite{sandvik10}. The velocity entering in  $\chi$ agrees with the velocity needed to fit the specific heat. 
The anomalous exponent $a \approx 0.22$ is in good agreement with an estimate based on a completely different analysis at $T=0$ \cite{lou09}, 
which suggests that the $T=0$ and $T>0$ exponents indeed are the same (which is unclear in the DQC theory, in which no anomalous $T>0$ exponent
has been discussed \cite{senthil04,kaul08}). The critical behavior does not fit the standard O($3$) picture with $S=1$ excitations \cite{chubukov94}, 
which we have investigated here in the context of a dimerized model. 

The agreement between the critical J-Q model and the non-interacting spinon gas is remarkable, considering that the spinons in the DQC theory are only 
marginally deconfined (with interactions mediated by the gauge field) \cite{senthil04}. Apparently, beyond their underlying role in determining the anomalous 
exponent $a$ in Eq.~(\ref{deltaspinon}), these interactions only have very small effects on the thermodynamics. A treatment similar to the spinon gas 
considered here has been applied to the $S=1/2$ Heisenberg chain (with the important difference that the spinons there obey Fermi statistics) 
\cite{mcrae}. Known results, including logarithmic corrections, were reproduced.

It would be useful to have an independent estimate of the spinon velocity. A velocity $c=2.4 \pm 0.3$ was extracted for the critical J-Q model
in \cite{melko08}, using a criterion for cubic space-time geometry in QMC simulations. Although the value is in good agreement with ours, it 
is unclear whether their method applies to spinons (while it should work for magnons). In future studies we will extract $c$ from 
imaginary-time dependent spin-spin correlations.

Our study lends support to the DQC scenario \cite{senthil04} for the N\'eel--VBS transition, although the phenomenological approach does not address 
the mechanism of deconfinement (only tests the consequences). Log corrections at $T>0$ should also have counterparts at $T=0$. Further work along these 
lines will hopefully explain, e.g., anomalous corrections to the spin stiffness of the J-Q model \cite{sandvik10} and its impurity response \cite{banerjee10}. 
An important missing link is how these corrections could arise from the CP$^{1}$ field theory of the DQC proposal \cite{senthil04}, i.e., whether this theory is complete in its current form or whether some ingredient is 
still missing. No log corrections were found in large-$N$ treatments of the CP$^{N-1}$ theory \cite{nogueira07,kaul08}, but it is 
possible that these corrections appear only for small $N$. A logarithmic enhancement of the susceptibility was found in a U($1$) gauge theory 
with fermions \cite{kim98}. In that case, there is also a correction to the specific heat, which makes the Wilson ratio non-divergent. The
spinon gas approach with Fermi statistics gives no log corrections and the Wilson ratio equals $0.0320$, which is identical (with $\mu=1/2$)
to the value in Ref.~\onlinecite{kaul08}.

We finally note that there are no indications of a first-order transition in the J-Q model (with previous claims \cite{jiang08,kuklov08} not 
supported by later results \cite{sandvik10,banerjee10}). As a matter of principle, however, extremely weak discontinuities cannot be ruled out 
based on numerical data alone (though the first-order scenario appears increasingly unlikely). What we have shown here is that, regardless of the 
ultimate nature of the transition, spinons are deconfined on length scales sufficiently large to have significant consequences for the 
thermodynamics.

{\it Acknowledgments}---We would like to thank Ribhu Kaul, Flavio Nogueira, Subir Sachdev and T. Senthil for useful discussions. 
AWS is supported by NSF Grant No.~DMR-0803510. VNK and OPS acknowledge support from the Condensed Matter Theory Visitors 
Program at Boston University.

\null

\end{document}